\documentclass[preprint,12pt,sort&compress]{elsarticlemod}
\usepackage{multirow}
\usepackage{overpic}
\usepackage{color}
\usepackage{rotating}

\usepackage{amssymb}

\journal{J. Environ. Radioact. 197 (2019) 39-47}

\begin{document}

\begin{frontmatter}



\title{Spatial Deconvolution of Aerial Radiometric Survey\\ and its application
  to the Fallout\\ from a Radiological Dispersal Device}


\author{Laurel E. Sinclair, Richard Fortin}

\address{Canadian Hazards Information Service, Natural Resources Canada,
  Ottawa, Ontario, Canada}

\begin{abstract}
Mapping radioactive contamination 
using aerial survey measurements is an area under active
investigation today.
The radiometric aerial survey technique has been extensively applied following reactor accidents
and also would provide a key tool for response to a malicious radiological or
nuclear incident.
Methods exist to calibrate the aerial survey system for quantification
 of the concentration of natural radionuclides, which can provide guidance.  However, these
methods have anticipated a spatial distribution of the source which is large in
comparison to the survey altitude.
In rapid emergency-response aerial surveys of areas of safety concern,
deposits of relatively small spatial extent may be expected.
The activity of such spatially restricted hot spots is underestimated
using the traditional methods.
We present here a spatial deconvolution method which can recover some of the
variation smoothed out by the averaging due to survey at altitude.
We show that the method can recover the true spatial distribution of
concentration of a synthetic source.
We then apply the method to real aerial survey data collected following
detonation of a radiological dispersal device.
The findings and implications of the deconvolution are then discussed by reference to a groundbased
truckborne survey over the same contamination.

\end{abstract}


\end{frontmatter}


\section{Introduction}
\label{sec:intro}
Aerial radiometric survey is a mature field.  Successful prospecting for
economically viable ore deposits using the radiation signal from naturally occuring
rocks stretches back decades~\cite{Grasty_1975}.
Survey systems composed of large volumes of NaI(Tl) scintillator gamma-ray detectors, as much as 20~L,
coupled with georeferenced position sensors (now making use of the global
navigation satellite system (GNSS)), record gamma energy spectra versus
position.  This information is later processed to produce maps of natural
potassium, uranium and thorium concentrations.  Standards exist to guide
practitioners in this area~\cite{IAEA_1991,IAEA_1995} and vast regions of the
earth have been covered~\cite{IAEA_2010}.
Practitioners have also developed methods to correct for terrain variation in
aerial survey~\cite{Schwarz1992,ISHIZAKI201782}.

The emphasis in aerial radiometric survey methods until recently has been on
development of techniques suitable for geologic sources, for which the
simplification of the source as an infinite and uniform sheet is reasonable in
comparison with the distance scales of the survey parameters (altitude, line
spacing).  The higher an aircraft flies, the more that far-away locations
contribute to the detected signal, relative to locations directly underneath
the aircraft~\cite{King_1912}.  This can have the
advantage of allowing for complete coverage in a more economical survey with
wider line spacing.  However, detection systems at higher altitude see a signal
which is effectively averaged over a larger area of the ground.  Anthropogenic
signals such as those resulting in case of a
reactor accident or malicious radiological dispersal could result in hot spots
the concentration of which would be underestimated if averaged over a larger area.

In this paper, we present a method to deconvolve an aerial radiometric survey
for spatial smearing.  This kind of problem, requiring inversion of a spatial
distribution, is encountered frequently in geophysical surveying.  
Geophysical spatial
inversion problems are typically underdetermined, and one way of dealing
with this has been to select only those solutions which are close to some
preconceived model~\cite{Parker_2015}.  An approach to spatial
deconvolution of airborne spectrometric data which relies on an analytical
model for the response function, and allows underdetermined problems, has been
published previously~\cite{Minty_2016}.  A related method for spatial
inversion to the approach presented here, but using an iterative inversion and
neglecting uncertainties, was published recently~\cite{Penny_2015}.  Other
groups are taking a similar approach to that advocated here, but applied to
the inversion of spectra rather than spatial
maps~\cite{Mattingly_2010, Hussein_2012}.

The method which will be presented here was applied to data obtained using
detectors aboard a manned helicopter.  Nevertheless, it is prudent to mention
the proliferation of work ongoing currently in aerial radiation detection
from unmanned aerial vehicle (UAV) systems.  The use of a UAV platform for 
aerial survey has facilitated the advance of 
aerial survey methodology.
A good review of recent publications can be found here~\cite{Connor_2016}.

The method which will be presented here involves simply a) determining the
influence of each independent region of the earth's surface on the measured spatial
distribution and then b) optimizing the weight coefficients of each region of
the surface to obtain the best reproduction of the measured map.  The number of
pixels of the solution can be chosen such that the problem is not underdetermined.  No
potentially biasing prior assumption about the underlying distribution is
necessary.  The method can handle complicated detector geometries as the
response functions are determined in simulation.  The method could easily be
extended to allow it to be applied when there is significant
terrain variation in the source such as would be the case in an urban
environment.  The method is stable under different starting
conditions, and naturally allows for propagation of uncertainties from the
measurement to the unfolded result.  

We demonstrate the application of the unfolding method by applying it first to
a synthetic data set.  This is compared with the known underlying distribution. 
We proceed to apply
the unfolding method to a real aerial survey measurement acquired following
detonation of a radiological dispersal device~\cite{Sinclair_RDD_2015}.

This spatial deconvolution method has been
presented previously at conferences by this
group~\cite{RDD_CTBT_2015,NSSMIC_RDD_2015}, however, this is the first full write-up.

\section{Methods}
\subsection{Unfolding method}
\label{sec:unfolding_method}
A measurement of surface activity concentration under the uniform infinite
plane assumption may be denoted $g^{\mbox{\scriptsize MEAS}}(x,y)$ where $x$
and $y$ represent easting and northing in geographic Cartesian coordinates.
We seek to determine the true underlying surface radioactivity concentration, $f(x,y)$.  $f(x,y)$ is related to  $g^{\mbox{\scriptsize MEAS}}(x,y)$ through
\begin{equation} 
g^{\mbox{\scriptsize MEAS}}(x,y) = S [f(x,y)],
\end{equation}
where $S$ represents the effect of the measurement system on $f(x,y)$.

We divide space into $N^{\mbox{\scriptsize PAR}}$ pixels $i$, and using Monte Carlo simulation, generate uniform radioactivity distributions in each pixel, $f_i(x,y)$.

The measurement system $S$ consists of the detection system as well as the air
and all other absorbing and scattering materials between the source and the
products of scattering, and the detectors.  It is represented in the Monte
Carlo simulation, and the emissions from the radioactive sources $f_i(x,y)$ are transported
through the system $S$ to obtain the template responses
$g_i(x,y)$, where
\begin{equation}
g_i(x,y) = S[f_i(x,y)].
\end{equation}

We let 
\begin{equation}
g(x,y) = \sum_{i=1}^{N^{\mbox{\tiny PAR}}} w_i g_i(x,y)
\end{equation}
and fit $g(x,y)$ to $g^{\mbox{\scriptsize MEAS}}(x,y)$ using a $\chi^2$ minimization~\cite{Minuit} to extract the weighting coefficients $w_i$.

To examine this $\chi^2$ function, let $g_j^{\mbox{\scriptsize MEAS}}(x,y)$ represent the $j$th measurement of the activity concentration $g^{\mbox{\scriptsize MEAS}}(x,y)$.  Then
\begin{equation}
\chi^2 = \sum_{j=1}^{N^{\mbox{\tiny MEAS}}}\frac{g_j^{\mbox{\scriptsize MEAS}}(x,y) - \sum_{i=1}^{N^{\mbox{\tiny PAR}}} w_i g_i(x,y)}{e_j^2}
\end{equation}
where there are $N^{\mbox{\scriptsize MEAS}}$ measurements $g_j^{\mbox{\scriptsize MEAS}}(x,y)$ in the problem each with uncertainty $e_j$.

The estimator of $f(x,y)$ is then the reconstructed radioactivity
concentration distribution $f^{\mbox{\scriptsize REC}}(x,y)$, where
\begin{equation}
f^{\mbox{\scriptsize REC}}(x,y) = \sum_{i=1}^{N^{\mbox{\tiny PAR}}} w_i f_i(x,y).
\end{equation}

We choose to require the problem to be oversampled.  That is, there is
everywhere a greater spatial density of measurements than of fit parameters
and $N^{\mbox{\scriptsize MEAS}} > N^{\mbox{\scriptsize PAR}}$.
Then, provided that the uncertainties $e_j$ in the denominator of the
$\chi^{2}$ function encompass all of the uncertainties of the problem, the
minimum $\chi^2$ value will be approximately equal to the number of degrees
of freedom of the problem.  
And the MINOS algorithm~\cite{Minuit} can be used to propagate the
$N^{\mbox{\scriptsize MEAS}}$ measurement uncertainties $e_j$ through the fit procedure
to calculate
the $N^{\mbox{\scriptsize PAR }}$ uncertainties 
$\delta w_i$ on the weighting parameters $w_i$.
In practice, there are irreconcilable nonstochastic uncertainties affecting the problem which
must be included in the
$e_j$ by application of a constant scaling factor
to bring $\chi^{2}$ per degree of freedom to one before the fit uncertainties
can be utilized.  These are due to the statistical uncertainties
in the template responses $g_i(x,y)$, and the finite pixellization of the problem.

Uncertainties for spatial deconvolution of fallout surveys can be expected to be asymmetric owing to the
boundary condition that the amount of deposition can not physically be less
than zero.  MINOS works by setting the positive and negative uncertainty for
each parameter to the amount the parameter has to vary in each direction such
that $\chi^{2}$ increases by one.  Thus MINOS naturally allows for assymetric
uncertainties and is particularly suited to uncertainty analysis in the
measurement of amount of radioactivity.

\subsection{Experimental method -- aerial survey}
\label{sec:exp_method}
The experimental methods to obtain the data to which we will apply the unfolding method have been described previously~\cite{Sinclair_RDD_2015}.  We will repeat only the most essential points here.
Three RDD detonations were conducted during the trial.  In the first, $\sim$31~GBq of \mbox{La-140} was dispersed explosively, with radioactive debris subsequently carried by wind as far as $\sim$~2~km from the site of the detonation.
Aerial gamma-ray spectrometric surveys were conducted using two
10.2~x~10.2~x~40.6~cm$^{3}$ NaI(Tl) crystals mounted exterior to a helicopter
in a skid-mounted cargo expansion basket.  GNSS antennae and inertial navigation and altimetry systems
were also installed in the basket, to determine location.
The system recorded a linearized 1024-channel gamma-ray energy spectrum over the domain 0~MeV to 3~MeV, tagged with the location information, once per second.
Post-acquisition, counts were selected from the spectra in an energy window of approximately four sigma in width around the 1.6~MeV \mbox{La-140} photopeak. 
These count rates were corrected for lag and dead time.  Count rates were
also corrected for variations in
flight altitude to the nominal flying height making use of the Shuttle Radar
Topography Mission~\cite{g.2007shuttle} digital terrain model
adjusted to the local elevation using a GNSS base station.
Backgrounds due to the radioactivity of the earth, the
atmosphere, the aircraft and its occupants, and cosmic rays, were all
subtracted.  The count rates were all corrected for radioactive decay, back to
the time of the blast.  A coefficient to convert the measurements from counts
per second to kBq/m$^2$, assuming an infinite and uniform source, was obtained
from experimentally validated Monte Carlo simulation.  
Finally, measurements of radioactivity concentration in kBq/m$^2$ for four
aerial surveys, two conducted after the first blast and one conducted after
each of the subsequent two blasts, were presented.

In this paper, we will discuss only the data recorded during the first aerial
survey after the first blast.
This survey was flown at a nominal 40~m flying height, with speed of
25~m/s and flight-line spacing of 50~m.

\subsection{Experimental method -- truckborne survey}
\label{sec:method_truck}

\subsubsection{Data collection -- truckborne survey}
Truckborne surveys were driven by criss-crossing the deposited fallout in an
extemporaneous pattern following
the first and third RDD blasts, restricting to the part of the fallout outside
of a 500~m x 500~m fenced hot zone~\cite{Green_RDD_2015,Marshall_thesis_2014}.  
The detection system was mounted in the bed of a pickup truck and consisted of four 10.2~x~10.2~x~40.6~cm$^{3}$
NaI(Tl) crystals oriented vertically in a self-shielding arrangement for
azimuthal direction measurement.
Truckborne data following the first RDD blast will be presented here for
comparison with the aerial survey data.  The truckborne data has not undergone
sufficient analysis for a full quantitative evaluation, but the shape will
nevertheless provide an interesting comparison for interpretation of the aerial survey data.

\subsubsection{Sensitivity calculation -- truckborne survey}

The sensitivity of the truckborne system to a \mbox{La-140} contamination on
the ground was determined using experimentally validated Monte Carlo
simulation.
In the simulations, the detector was placed with the centre of its sensitive
volume at a height of 1.2~m above ground.  
The NaI(Tl) crystals and their housing were represented in the simulation in 
their vertical arrangement. 
The $\sim 5000$~kg of mass of the pick-up
truck carrying the detector was represented by simple blocks of steel.

Sensitivity validation data was collected by placing a \mbox{La-140} source of
known emission rate at fixed locations around the truck and detector system.
The dead material was adjusted in size and position in the model until an acceptable agreement with the
sensitivity validation data was obtained.
The engine and other materials at the front of the truck effectively block
radiation coming from that direction, and most of the detected counts arise
from radiation originating to the side and rear of the truck where there is
comparatively little material.
The uncertainty in the estimation of the sensitivity of the truckborne
detector obtained in this manner is large, about 20\% to 30\%.
This level of accuracy is sufficient for illumination of the value of the spatial deconvolution applied to the aerial
data by comparison with data collected from a ground-based system.

\begin{figure}[h]
   \begin{center}
   \begin{tabular}{c}
   \begin{overpic}[trim = .1cm .1cm .1cm .1cm, clip = true, height=4.7cm]{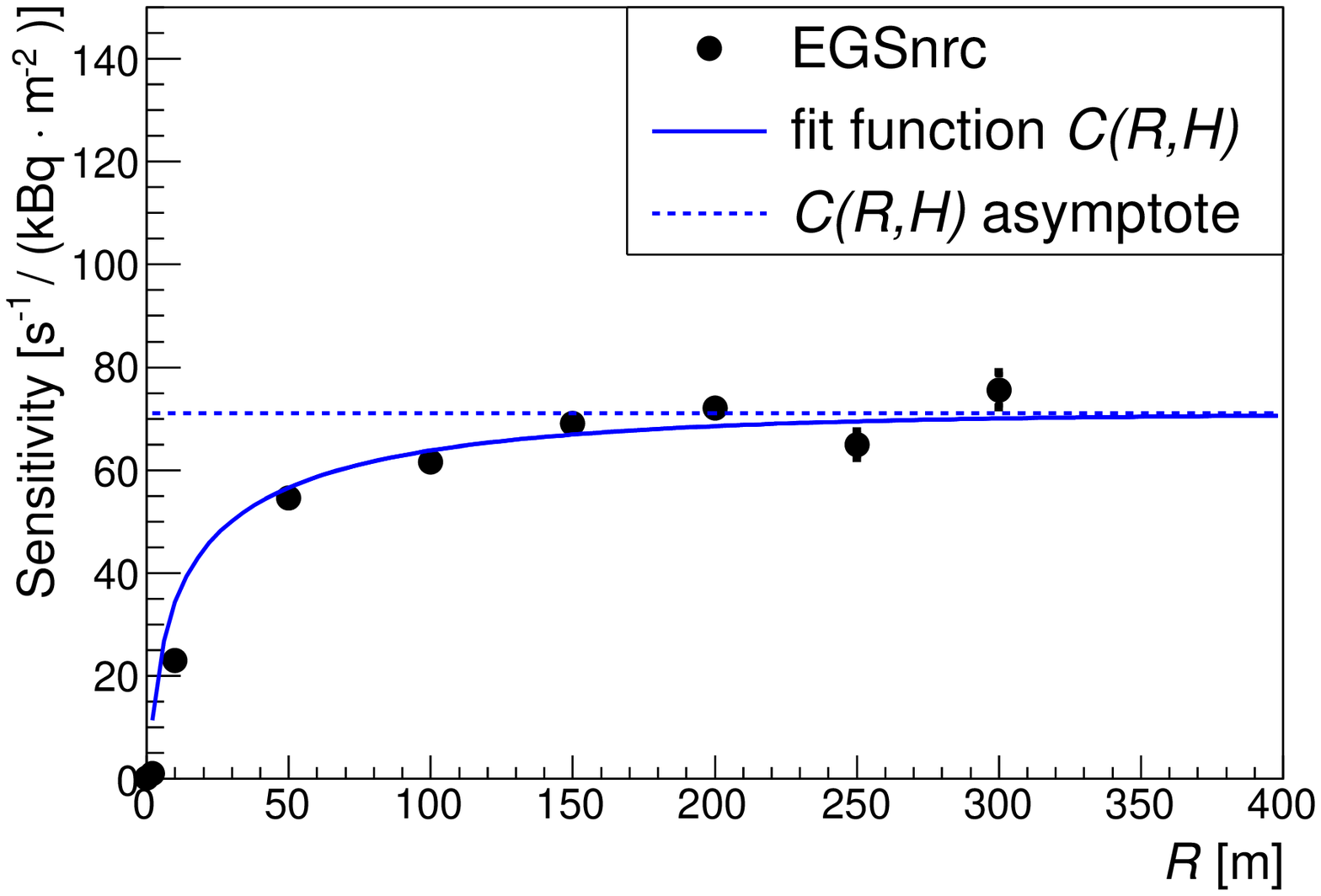}\put(18,52){\textcolor{black}{a)}}\end{overpic}
   \hspace*{-.1cm}
   \begin{overpic}[trim = .1cm .1cm .1cm .1cm, clip = true, height=4.7cm]{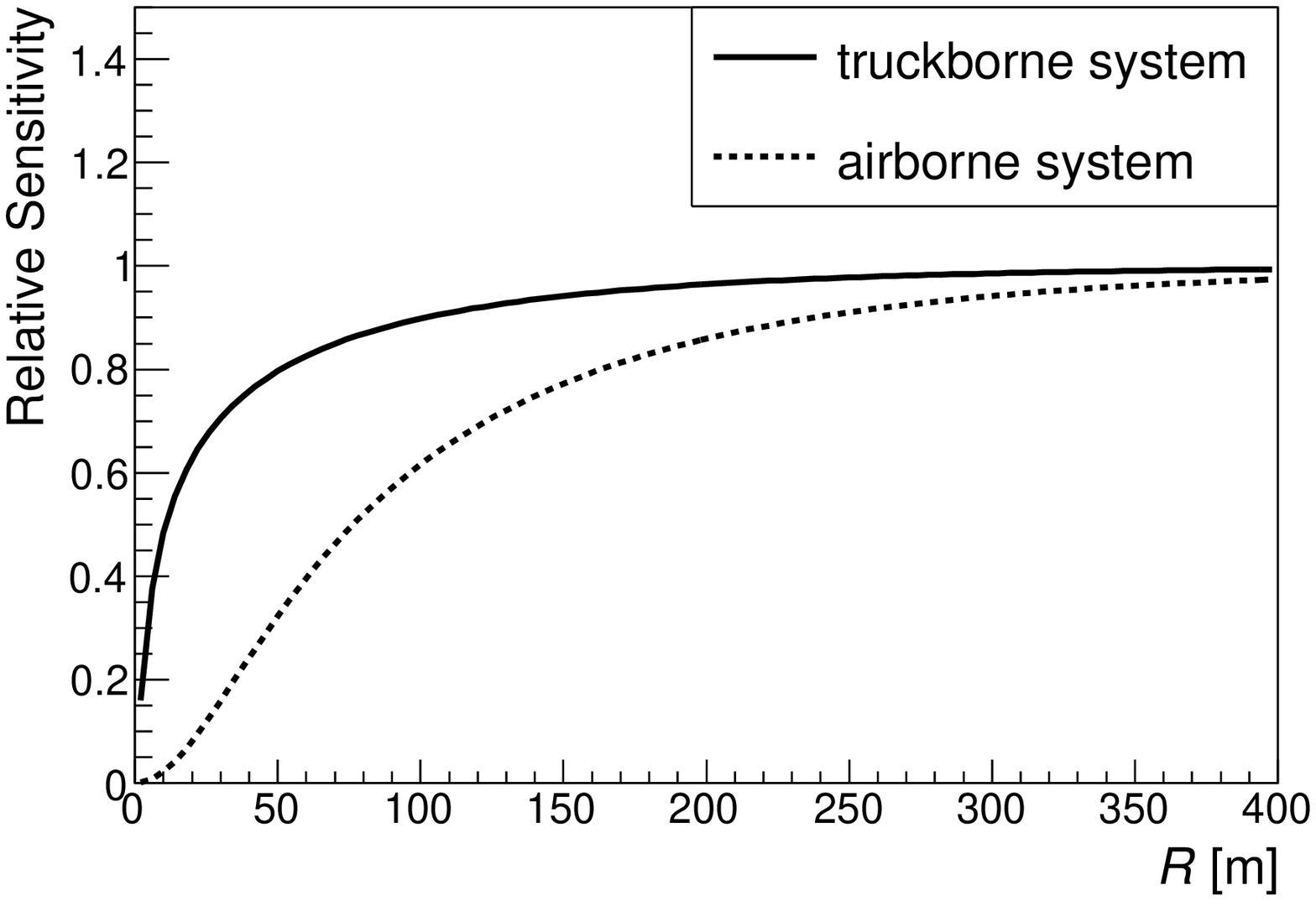}\put(18,52){\textcolor{black}{b)}}\end{overpic}
   \end{tabular}
   \end{center}
  \caption
  { \label{fig:truck_sensitivity} 
a) Sensitivity of the truckborne survey system to a disc-shaped distribution of isotropic emitters of the \mbox{La-140} gamma spectrum, as a function of disk radius.  Black dots show EGSnrc prediction.  Solid curve shows fit of the expression $C(R,H)$ to the synthetic data.  The dashed line shows the asymptote of the fit curve and represents the sensitivity to a uniform and infinite sheet source.
b) Comparison of the shapes of the sensitivity curves for the aerial and truckborne survey systems.  The dashed line shows the sensitivity to a disc source relative to the sensitivity to an infinite sheet as a function of disc radius for an aerial survey system at an altitude of 40~m~\cite{Sinclair_RDD_2015}.  The solid line shows the equivalent relative sensitivity curve for the truckborne survey system.
}
\end{figure}

Fig.~\ref{fig:truck_sensitivity}~a) shows the number of energy deposits
registered in the detector per second as a function of the radius $R$ of a
disc-shaped source centered beneath the truckborne detector on the ground as
determined by the simulation.
The expression for the flux, $\Phi(R,H)$, due to a surface activity concentration $S_0$ at a point an elevation $H$ above a disc-shaped source of radius $R$ can be readily calculated~\cite{King_1912},
\begin{equation}
  \Phi(R,H) = \frac{S_0}{2} (E_1(\mu_{\mbox{\scriptsize a}}H) - E_1(\mu_{\mbox{\scriptsize a}}\sqrt{H^2+R^2})),
\label{eqn:flux_vs_R}
\end{equation}
where E$_1$ is the exponential integral and $\mu_{\mbox{\scriptsize a}}$ is the linear attenuation coefficient for gamma rays in 
air.
To determine the asymptotic sensitivity, we formed a function for the detected
counts as a function of the source radius, 
\begin{equation}
   C(R,H) = \epsilon \Phi(R,H),
\label{eqn:count_rate}
\end{equation}
and fit the expression for $C(R,H)$ to the synthetic data to obtain the
detection efficiency $\epsilon$.
The fit result is shown as the solid curve in
Fig.~\ref{fig:truck_sensitivity}a) and the asymptotic sensitivity, shown by
the dashed line, is $\sim 71$~s$^{-1}$/(kBq/m$^2$).
As mentioned, the uncertainty on this sensitivity is large due to the lack of
detailed representation of the shielding material of the truck in the
simulation.
Nevertheless, the shape of the sensitivity curve is of value, as is the shape
of the profile
of counts measured with the truckborne system as it traversed the deposited plume.

\subsubsection{Comparison of aerial and truckborne sensitivity curves}
Fig.~\ref{fig:truck_sensitivity}b) shows a comparison of the shapes of the
sensitivity curves of the ground-based and aerial systems.  Despite the tall
narrow shape of its detectors, which would tend to increase sensitivity to incoming radiation
from the sides, for the truckborne system at ``altitude'' $H=1.2$~m, a greater
percentage of detected gamma rays originate close to the point directly
underneath the detector as compared to the airborne
system at $H=40$~m.
This leads to superior spatial precision in the results of truckborne survey.

\subsection{Aerial survey template response determination through Monte Carlo simulation}
\label{sec:simulation}
\begin{figure}[h]
   \begin{center}
   \begin{tabular}{c}
   \begin{overpic}[trim=0cm -5cm 0cm 3cm, clip = true, height=5.1cm]{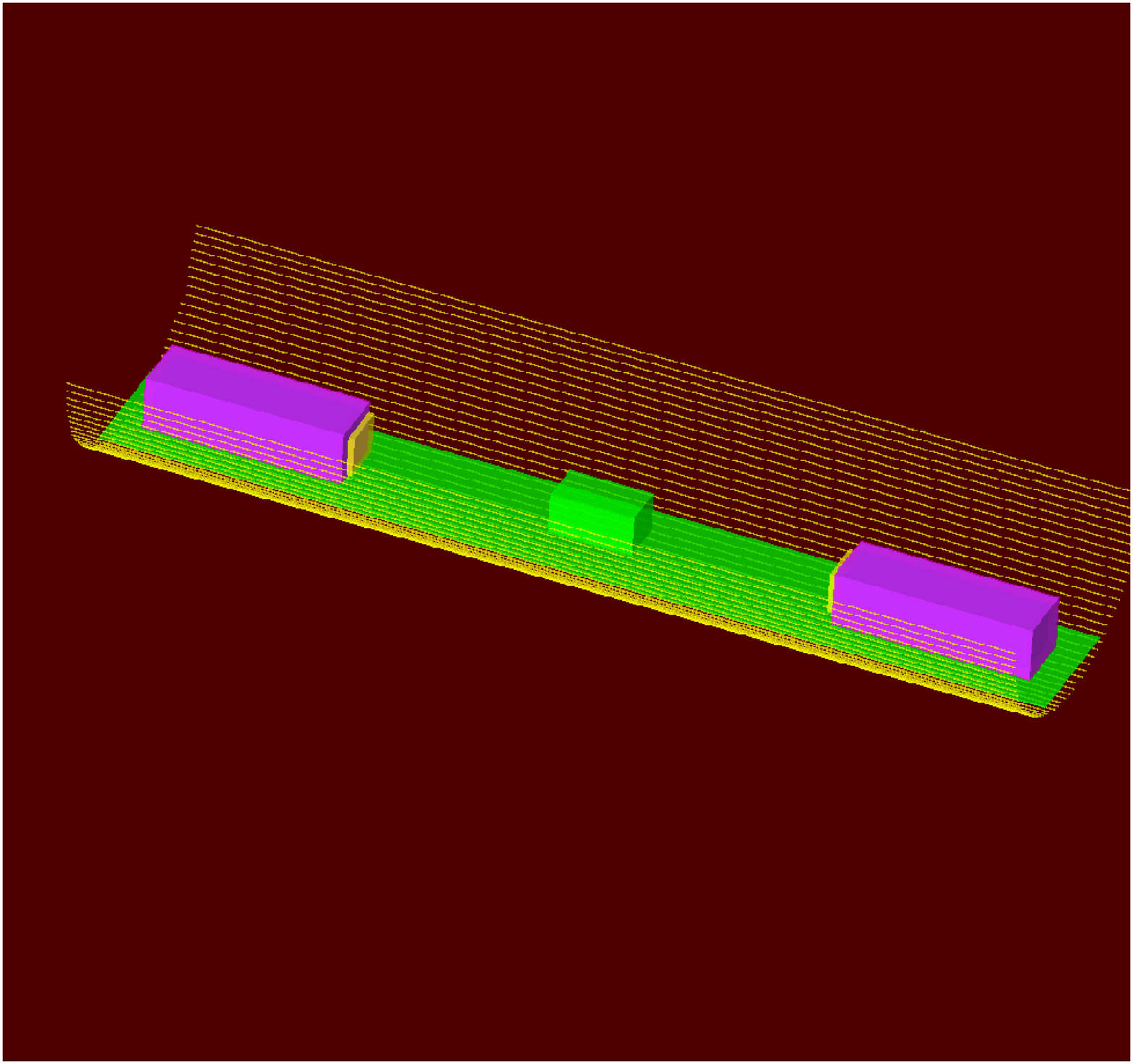}\put(5,89){\textcolor{white}{a)}}\end{overpic}
   \hspace{1.4cm}
   \begin{overpic}[height=5.5cm]{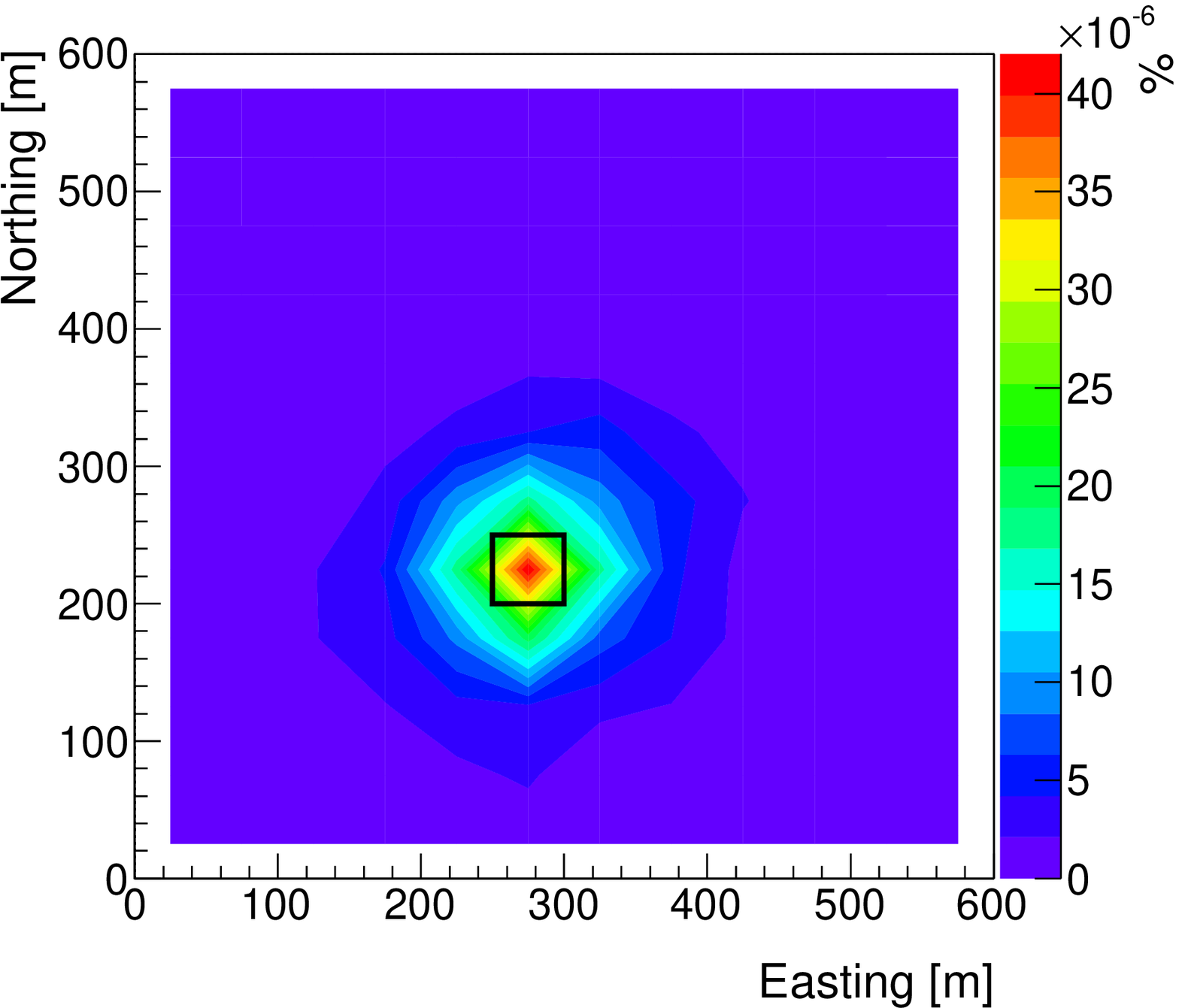}\put(18,70){\textcolor{white}{b)}}\end{overpic}\\
   \end{tabular}
   \end{center}
  \caption
  { \label{fig:egspp} 
a) The aerial survey system as modelled in EGSnrc.  The two
10.2~x~10.2~x~40.6~cm$^{3}$ NaI(Tl) crystals
are represented in their housing, shown in purple, with steel blocks at the
ends representing the dead material of the PMTs and readout electronics.  The
two crystals are mounted lengthwise with their centres 152.4~cm apart on an aluminium plate which is 
represented with mass and dimensions according to the engineering drawing.
Auxilliary instrumentation is represented by an
aluminium block in the centre of the basket, of the summed instrument mass.
This dead material is shown in green in the figure.  The aluminum plate on
which the detectors and auxiliary equipment is mounted is itself attached to a
basket which is mounted to
the skids on the outside of the helicopter.  Dead material of the basket is represented by 51 thin aluminum bars running the length of the basket, in a
semicircle around the basket long axis, shown in gold.  Uncertainty on the
sensitivities due to misrepresentation of the system in the model has been
estimated to be approximately 12\%.
b) Response function, $g_{53}(x,y)$ to the pixel source $f_{53}(x,y)$,
normalized to the number of generated events, where the $x$-axis shows Easting
and the $y$-axis shows Northing.  The true spatial extent of pixel 53 is indicated by the black square.}
\end{figure}

The radiation transport model EGSnrc~\cite{EGSnrc1,EGSnrc2} was used to
generate the individual uniform pixel sources $f_i(x,y)$ and to propagate the
generated gamma rays and their interaction products through air and into the detection volume to create the
template responses $g_i(x,y)$.  For the solutions presented herein, the
simulation geometry represented the experimental setup during the first RDD
trial~\cite{Sinclair_RDD_2015}.  The actual aerial survey system was described and shown
in photographs in the previous publication and briefly reiterated in 
Sect.~\ref{sec:exp_method}.
The model of that system in EGSnrc is shown in
Fig.~\ref{fig:egspp}~a).  The simulated gamma detection system included the
two NaI(Tl) crystals, as well as their aluminum cladding, and felt, foam and
carbon fibre enclosure.  The exterior-mounted basket containing the detectors
was modelled in a simplified manner with 51 3~mm~x~1.5~mm bars of aluminum,
representing the basket strands, running the length of the basket in a
semicircle around its long axis.  Dead materials due to the photo-multiplier
tube readout of the crystals and associated electronics, as well as the
altimeter and GNSS receivers, were modelled as simple blocks of metal of the
appropriate overall mass.  The ground was represented as perfectly flat with
the detection system at a height of 40~m.  
The model was validated experimentally using data from point sources of known
emission rate placed at
known locations around the actual detector.  The uncertainty associated with approximation in the representation of the
measurement system in the model was evaluated by variation of the arrangement
of dead material in the model, by variation of the detector's position,
altitude and
attitude and by variation of the detector's energy resolution within
reasonable limits.  This uncertainty was evaluated to be about 12\% on the 
activity
  concentration and it was included the overall systematic
uncertainty quoted in the  publication~\cite{Sinclair_RDD_2015}.

The simulated pixel sources, $f_i(x,y)$, consisted of uniform distributions of
isotropic emitters of the \mbox{La-140} spectrum of gamma rays, including all
emission energies above 0.05\% emission probability~\cite{TabRad_v1}.  Gammas
were emitted into the full 4$\pi$ solid angle.  Each pixel source $f_i(x,y)$
was square, and 50~m on a side.  Note that with the survey parameters
mentioned previously, we have one spectrum recorded approximately every
25~m~x~50~m in the real data.  Thus, the number of measurements in the data is
higher than the number of fit parameters in the simulation, the problem is over-determined, and we can expect reasonably rapid convergence of the method to a solution which is stable under different starting conditions.

The entire region to be unfolded measured 1.5~km~x~1.5~km.  To speed up
convergence of the fitting, we chose to parallelize the computation, breaking
the problem up into individual tiles, each 500~m~x~500~m.  Only eight of these
tiles were necessary to cover the area over which the radioactivity actually
settled.  To knit the tiles together, we extended the fit into a border of
pixel sources 50~m wide, surrounding each tile.  Thus the fit area
corresponding to each tile was actually 600~m~x~600~m.  If, after background
subtractions, fewer than one count corresponding to a \mbox{La-140} energy
deposit was measured in the detection system in the region of space
corresponding to one fit parameter $w_i$, then that fit parameter was assigned
a value of zero and left out of the minimization.  Thus, 144 or fewer
fit parameters $w_i$ were determined from the inversion of each tile.
To merge the tiles after the individual inversions, the weighting parameters of the border pixels were simply ignored, and the central 500~m~x~500~m areas of the tiles were placed next to each other.

The simulation included one detection system for each of the 144 pixel sources, centered
laterally within the pixel, at 40~m height.  
Here we have used multiple detection systems at different places at one time to
represent the real world in which one detection system was in different places
at different times.  Given the 40~m height of the detection systems above the
source, the 50~m spacing between them and the requirement that the deposited
energy lie in the highest-energy photopeak of the source, the error in this
approximation, which would come from a single emitted gamma ray depositing energy in two
different detection systems, is negligible.

The volume of the air in which the gamma rays and electrons were tracked in each
tile extended 1.2~km in easting (or $x$) and northing (or $y$), and 600~m up.
A 5~m-thick concrete slab underneath the emitters and filling the lateral
dimensions of the simulated volume, represented the earth.

The simulation included all physical interactions of the emitted gammas and of
the gammas, electrons and positrons resulting from those interactions.  Scattering
and absorption in the air and ``earth'' of the simulated volume, and in the dead material
of the basket system and housing of the NaI(Tl) crystals, was included.  All
processes leading to energy deposit within the NaI(Tl) crystal from either the
direct gamma rays, or the products of scattering, were included.  Energy
deposits in the NaI(Tl) were then smeared to create simulated spectra,
according to the energy resolutions of the crystals determined during the experiment.

Fig.~\ref{fig:egspp}~b) shows the response, $g_{53}(x,y)$, of the 144
detection systems of one tile, as a percentage of the number of events
generated in a single one of the pixel sources, $f_{53}(x,y)$, where this
happens to correspond to the pixel numbered ``53''.  The extent of the source
is indicated by the black square.  Note how the measured response extends much
more broadly in space.  This is the spatial smearing which will be corrected
by the unfolding.

\section{Results}

\subsection{Results obtained by application of the method to synthetic data}
\begin{figure}[h]
   \begin{center}
   \begin{tabular}{c}
   \begin{overpic}[trim = .1cm .1cm .1cm .1cm, clip = true, height=6.5cm, angle=270]{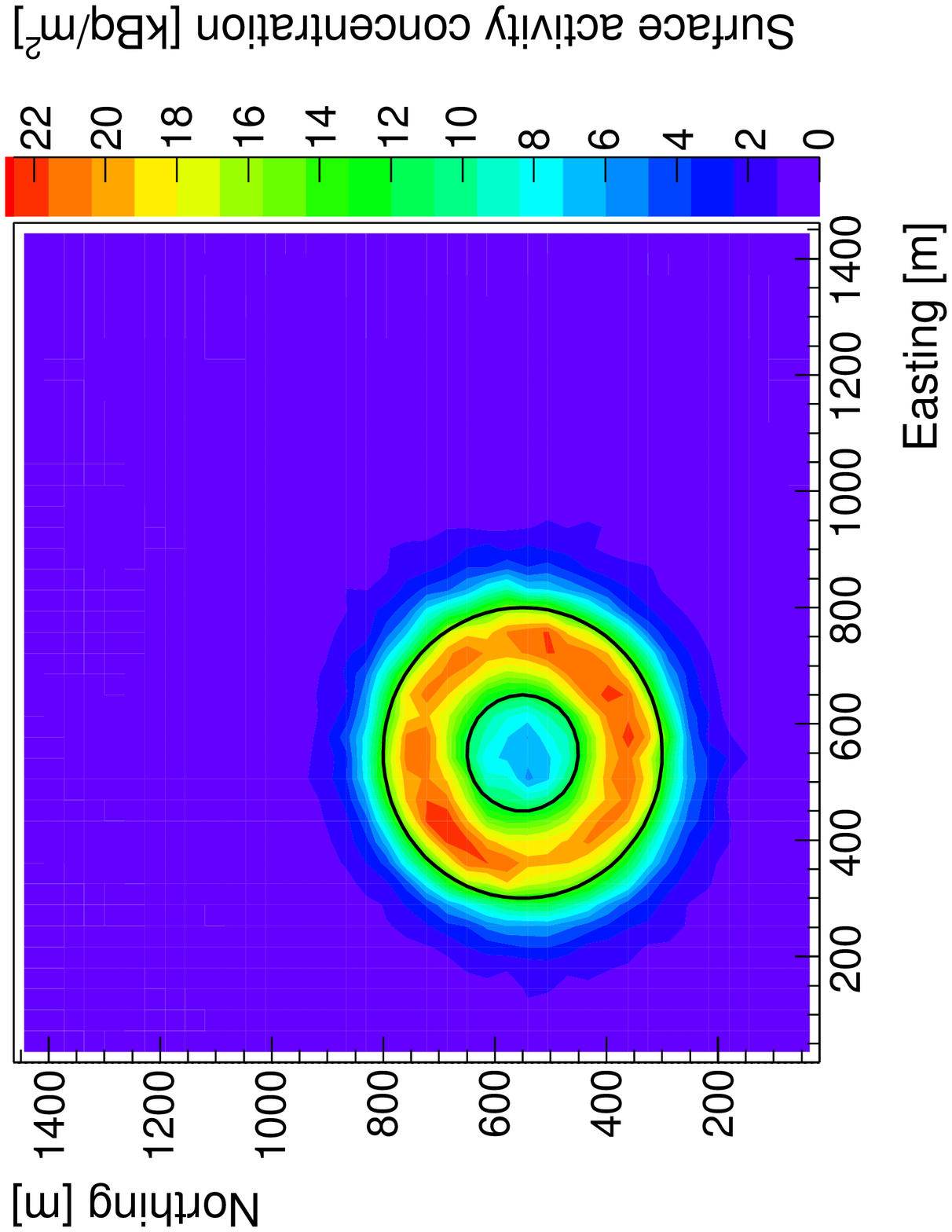}\put(18,70){\textcolor{white}{a)}}\end{overpic}
   \hspace{.4cm}
   \begin{overpic}[trim = .1cm .1cm .1cm .1cm, clip = true, height=6.5cm, angle=270]{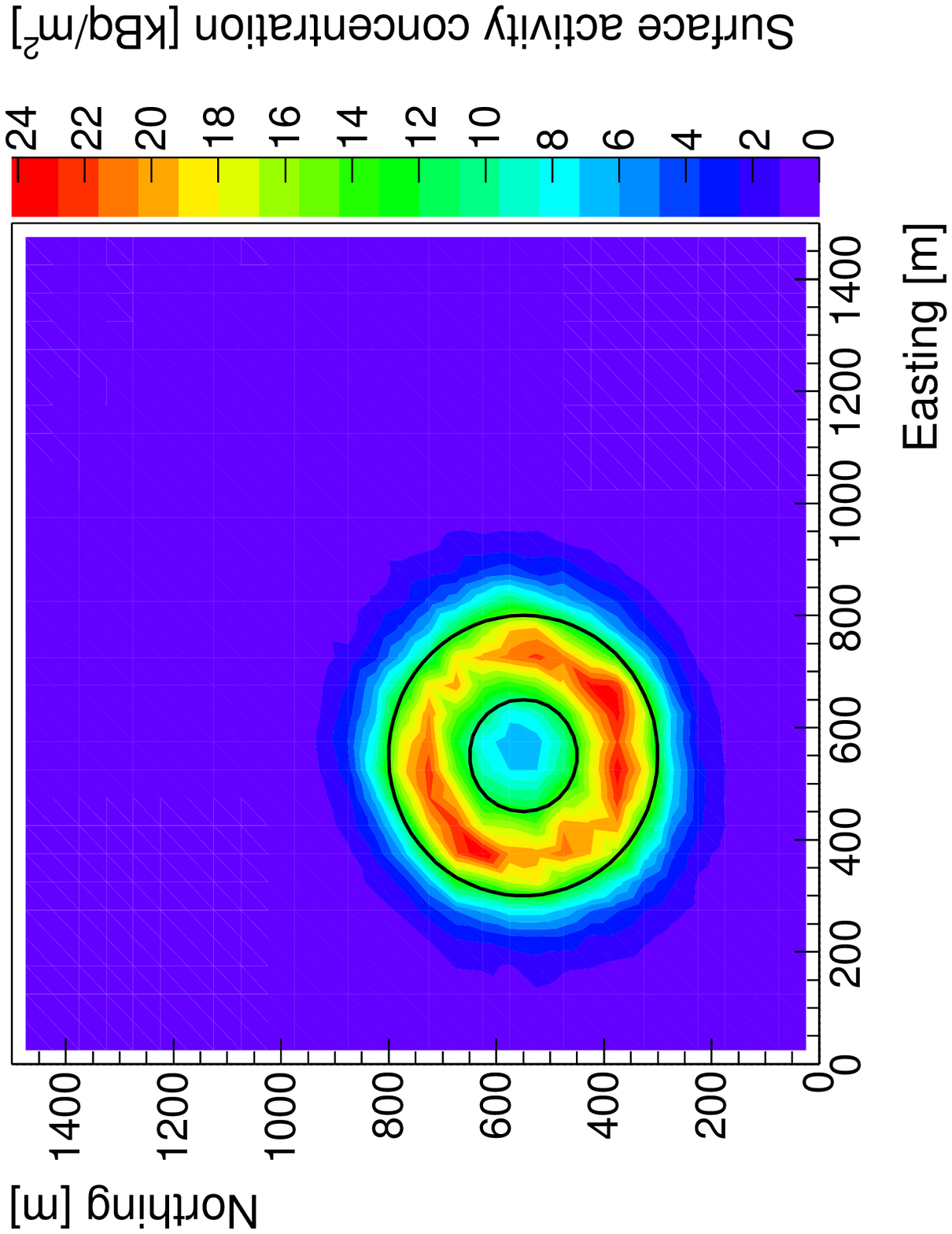}\put(18,70){\textcolor{white}{b)}}\end{overpic}\\
   \end{tabular}
   \end{center}
  \caption
  { \label{fig:annulus_data_and_fit} 
Synthetic aerial survey and fit results, where the $x$
axis is Easting, and the $y$ axis is Northing.  The source was of concentration 
28.4~kBq/m$^2$ and
annular with inner radius 100~m and outer radius 250~m, centered at
(550~m, 550~m), as indicated by the area between the black circles.
a) Synthetic aerial survey measurement result.
b) Result of fit of template responses $g_i(x,y)$ to the synthetic aerial survey.}
\end{figure}

We begin by applying the spatial inversion method to a synthetic data set for
which we know the underlying distribution $f(x,y)$.  
An annular region consisting of the area between two circles of radius 100~m
and 250~m, centered at (550~m, 550~m) was uniformly populated with
10~$\cdot$~10$^{9}$ emitters of the \mbox{La-140} gamma sectrum.  Considering the
branching ratios for the \mbox{La-140} gamma emissions, this amounts to a source
concentration of 28.4~kBq/m$^2$.  The annular region
is indicated by the black outlines in Fig.'s~\ref{fig:annulus_data_and_fit}~a)
and~b).

Generation of the synthetic dataset makes use of the
identical detector simulations as used in generating the template 
responses $g_i(x,y)$ as described in Section~\ref{sec:simulation}, however the detection systems were more narrowly spaced
in the synthetic dataset.
Detection systems for the synthetic dataset were located every 20~m in $x$ and
$y$ such that there were 225 detection systems in total in each 600~m~x~600~m
tile.

The template sources $f_i(x,y)$ and the template responses $g_i(x,y)$ utilized
in the inversion are the same as will be used for the real data and are as
described in Section~\ref{sec:simulation}.
Thus, the synthetic data measurement density of 20~m~x~20~m exceeds the
density of the parametrization, 25~m~x~25~m, and the problem
is overdetermined, as desired.

Fig.~\ref{fig:annulus_data_and_fit}~a) shows the synthetic aerial survey
measurement.  The result is broader than the underlying true source
distribution.  Contamination appears to extend outside of the known true
underlying borders.  The central area of the annulus appears to be filled with
significant contamination.  The average concentration of contamination
reconstructed within the annular region is lower than the known true concentration.  

Fig.~\ref{fig:annulus_data_and_fit}~b) shows the result of the fit of the
template measured activity distributions to the synthetic aerial survey
measurement.  The colour scale used in Fig.~\ref{fig:annulus_data_and_fit}~b) is
the same as that of Fig.~\ref{fig:annulus_data_and_fit}~a).
The first observation to note is that the tile knitting procedure
apparently works well.  The synthetic data stretches over six of the
overlapping 600~m x
600~m simulation tiles.  After knitting of the inverted result into adjacent 
non-overlapping
500~m~x~500~m tiles, there is seamless matching of the reconstructed activity
concentration at the tile borders.
Also to note is that within the limitations of the somewhat coarse pixellization of
the problem, the survey is well reproduced by the fit.  
The tendency of the
measurement to extend outside of the bounds of the true source is reproduced,
as is the tendency of the measurement to underestimate the magnitude of the
concentration within the source boundary.

\begin{figure}[h]
   \begin{center}
   \begin{tabular}{c}
   \begin{overpic}[trim = .1cm .1cm .1cm .1cm, clip = true, height=6.5cm, angle=270]{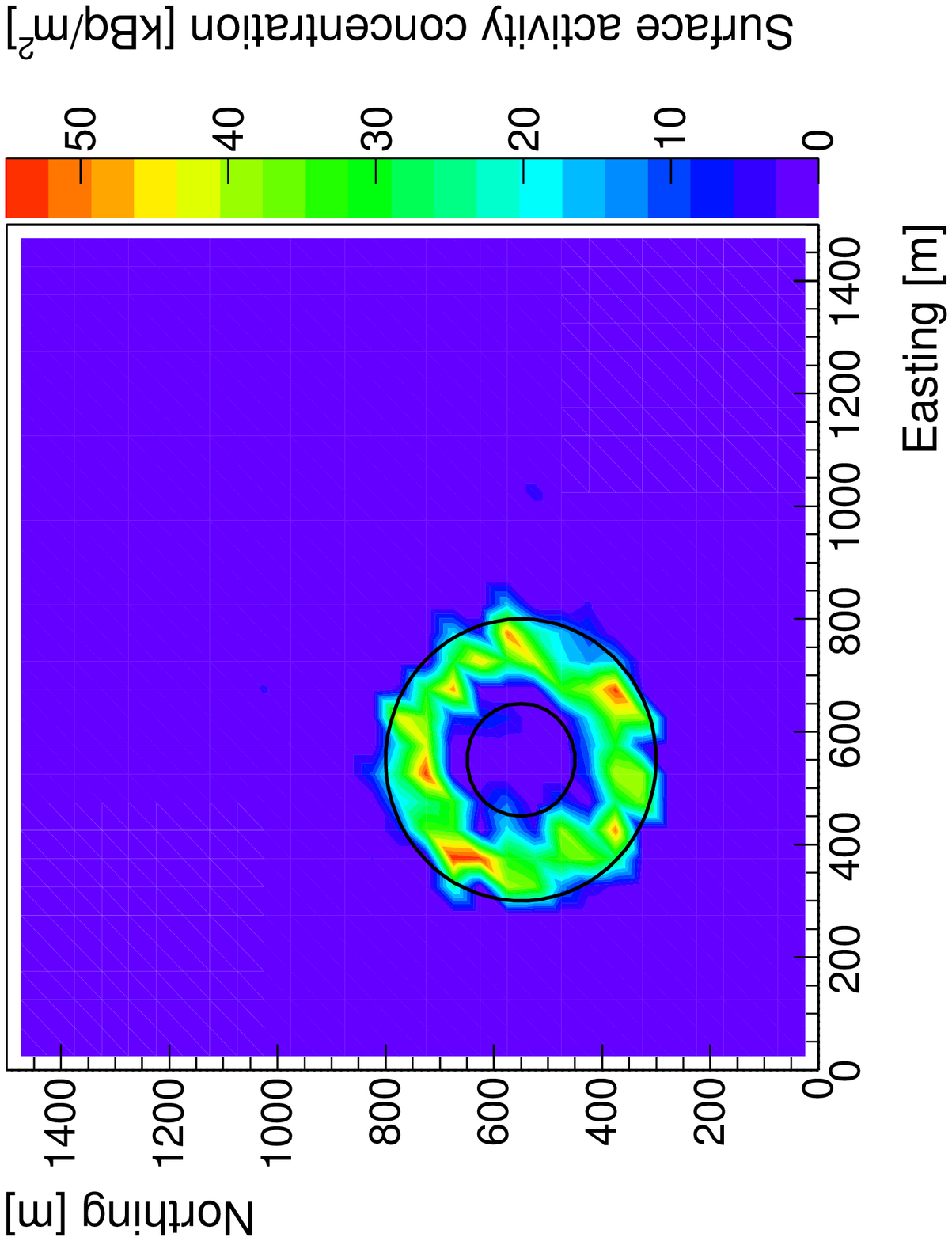}\put(18,65){\textcolor{white}{a)}}\end{overpic}\\
   \begin{overpic}[trim = .1cm .1cm .1cm .1cm, clip = true, height=6.5cm, angle=270]{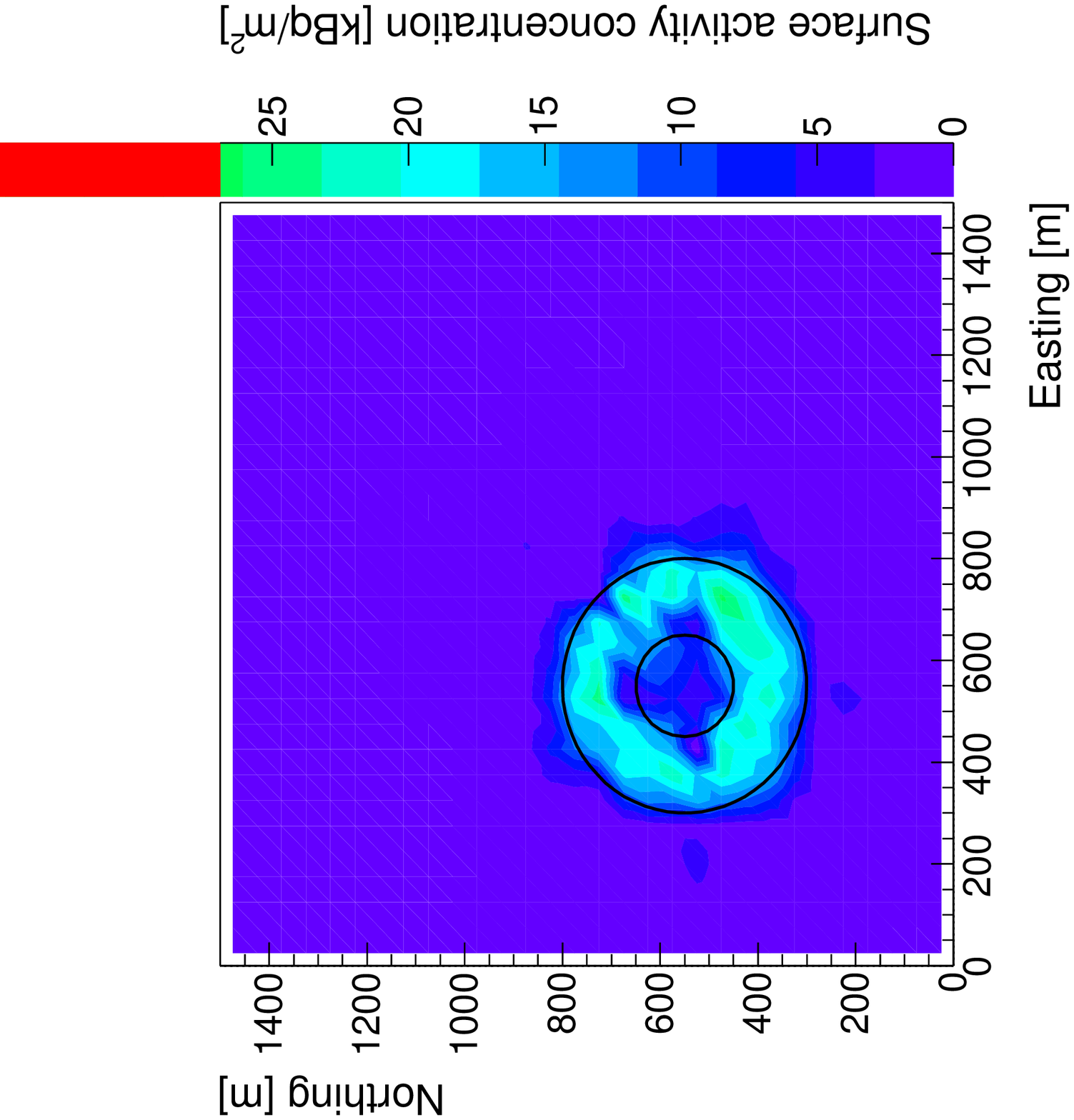}\put(18,65){\textcolor{white}{b)}}\end{overpic}
   \hspace{.4cm}
   \begin{overpic}[trim = .1cm .1cm .1cm .1cm, clip = true, height=6.5cm, angle=270]{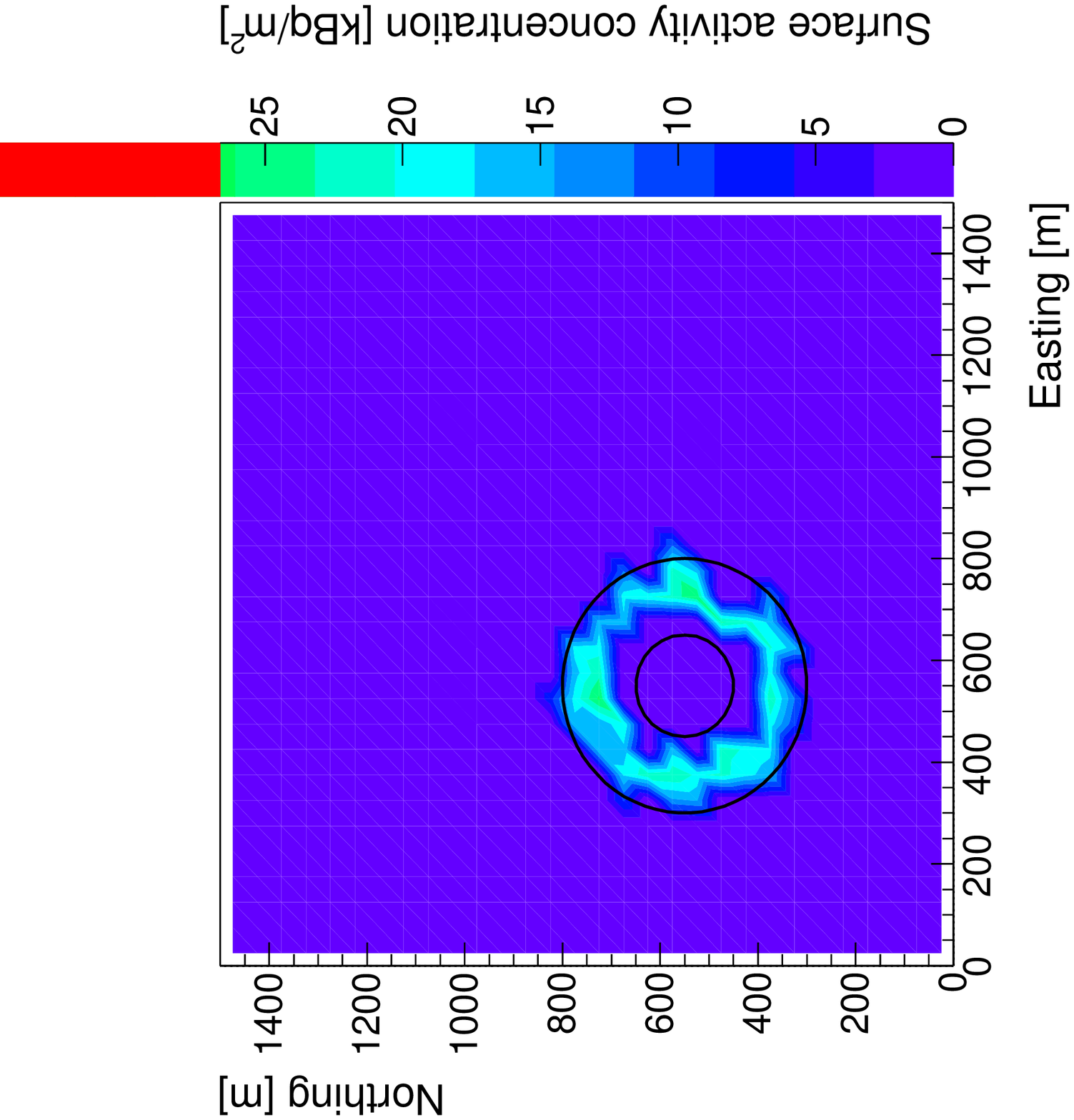}\put(18,65){\textcolor{white}{c)}}\end{overpic}\\
   \end{tabular}
   \end{center}
  \caption
  { \label{fig:annulus_inverted} 
Spatially deconvolved synthetic aerial survey result, $x$ axis is Easting and
$y$ axis is Northing.  Black circles indicate
the true annular source distribution.
a) The spatially deconvolved measurement.
b) Positive statistical uncertainty on the spatially deconvolved measurement.
c) Negative statistical uncertainty on the spatially deconvolved measurement.}
\end{figure}

The reconstruction of the true underlying source distribution for the
synthetic data is presented in Fig.~\ref{fig:annulus_inverted}.
As shown in Fig.~\ref{fig:annulus_inverted}~a), the inversion process results
in a reconstructed source distribution which is higher in magnitude and closer
to the true activity concentration than the
initial survey measurement.  The boundaries of the actual source are 
better reproduced after inversion, in particular the absence of radioactivity
in the centre of the annulus is recovered.

A major advantage of the spatial
deconvolution method presented here is that statistical uncertainties affecting
the measurement may be propagated through the minimization procedure by the MINOS algorithm as described
in Sect.~\ref{sec:unfolding_method}.
 A map
showing the one-sigma positive statistical uncertainty on the reconstructed surface activity
concentration is shown in
Fig.~\ref{fig:annulus_inverted}~b) and the corresponding negative statistical uncertainty
is shown in Fig.~\ref{fig:annulus_inverted}~c) where the same colour scale is
used for the uncertainties as for the measurement shown in 
Fig.~\ref{fig:annulus_inverted}~a).
Considering the uncertainties, the ability of the method to reconstruct the
true underlying activity concentration is good.  The reconstructed activity
concentration magnitude is in agreement with the known true activity
concentration within uncertainties in most places.  For example, near the
inner boundary of the annulus where the reconstructed concentration is low
compared to the known true concentration, further negative movement of the
result is not allowed by the negative uncertainty.  The positive uncertainty exceeds the
negative uncertainty in magnitude, and does allow for a positive movement of the
reconstructed value.

\subsection{Real data collected following detonation of a radiological
  dispersal device}
\subsubsection{Spatial deconvolution of aerial survey data}

\begin{figure}[h]
   \begin{center}
   \begin{tabular}{c}
   \begin{overpic}[trim = .1cm .1cm .1cm .1cm, clip = true, height=6.5cm, angle=270]{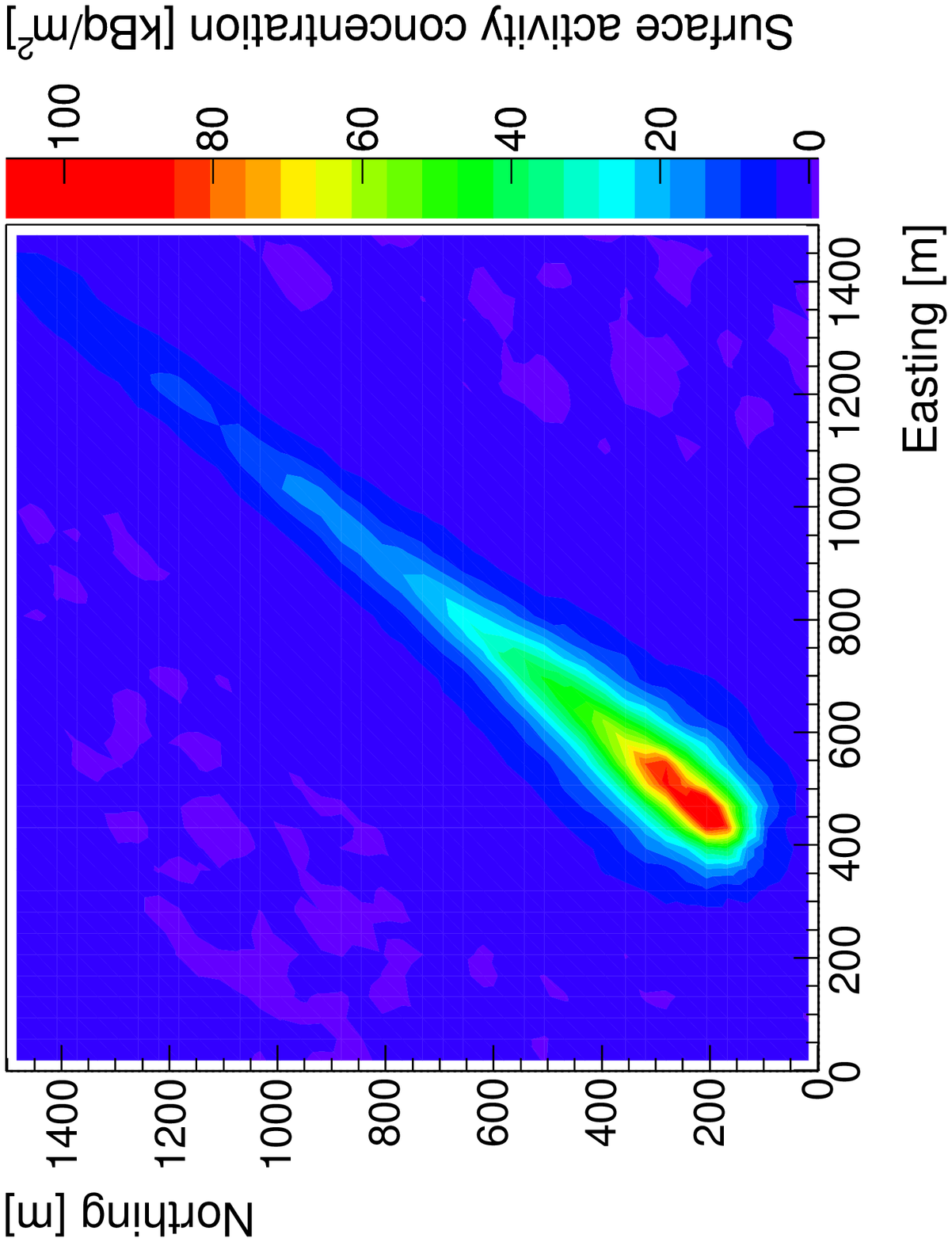}\put(18,70){\textcolor{white}{a)}}\end{overpic}
   \hspace{.4cm}
   \begin{overpic}[trim = .1cm .1cm .1cm .1cm, clip = true, height=6.5cm, angle=270]{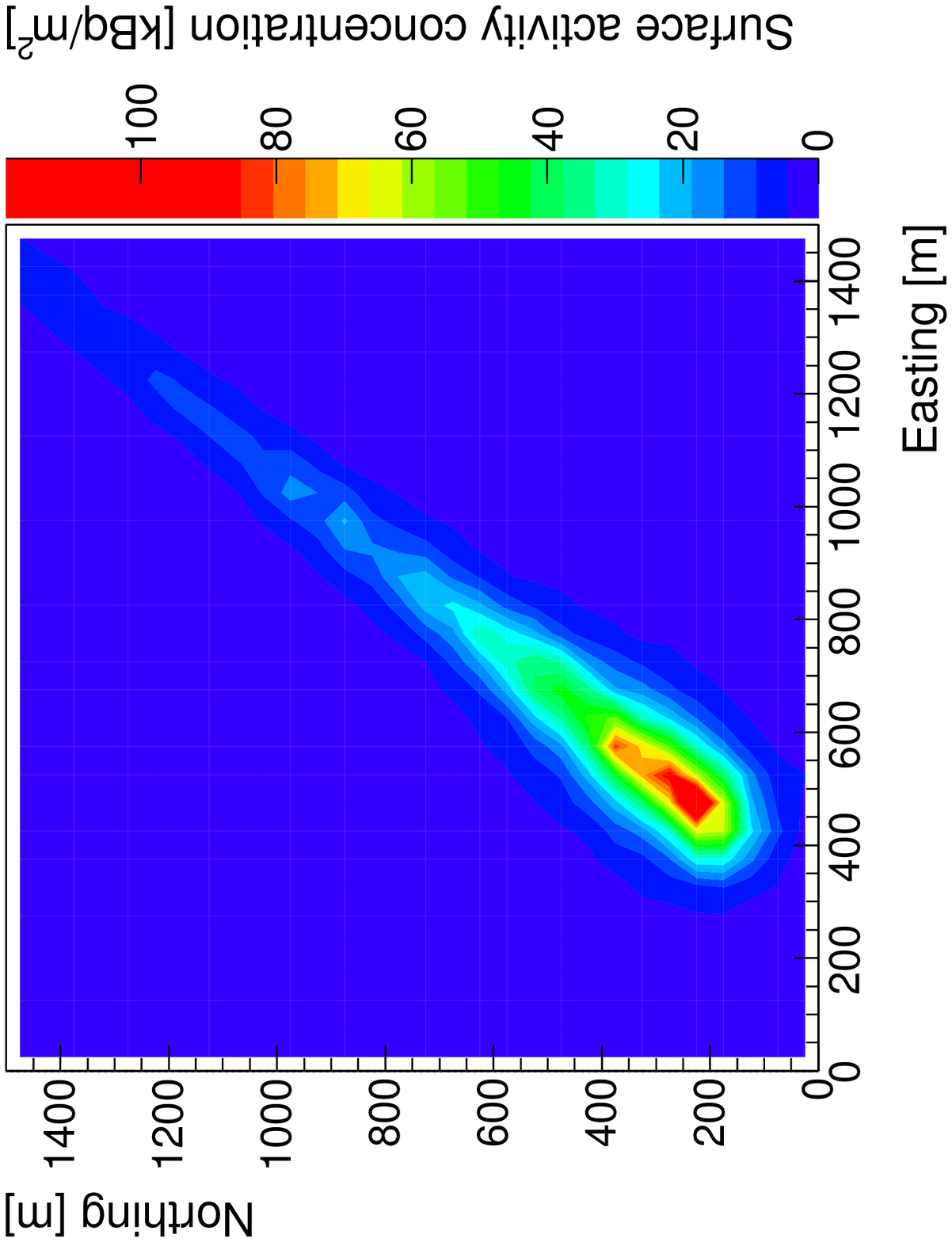}\put(18,70){\textcolor{white}{b)}}\end{overpic}\\
   \end{tabular}
   \end{center}
  \caption
  { \label{fig:data_and_fit} 
a) Aerial survey measurement of the distribution of fallout following detonation
  of the radiological dispersal device.
b) Result of fit of template histograms to the aerial survey measurement.}
\end{figure}

The aerial survey measurement from the first RDD trial is shown in Fig.~\ref{fig:data_and_fit}~a).
This result has been published previously~\cite{Sinclair_RDD_2015} and the methods to
arrive at the result were recapitulated here in Sect.~\ref{sec:exp_method}.
We observe an area of activity concentration exceeding 100~kBq/m$^2$ near
ground zero of the detonation, with a long deposited plume of maximum width
of 300~m to 400~m, and significant measured radioactivity extending over a
distance of about 2~km.
The total systematic uncertainty affecting this measurement was determined to
be around 12\% and the statistical uncertainty peaks at 6~kBq/m$^2$.

Fig.~\ref{fig:data_and_fit}~b) shows the result of the $\chi^2$ fit of
the weighting coefficients of the template response functions to the
measurement of Fig.~\ref{fig:data_and_fit}~a).  The colour scales of 
Fig.'s~\ref{fig:data_and_fit}~a) and~b) have been chosen to be equal.  
(This colour
scale has in fact been set to the optimal colour scale for the data from a
truckborne survey which will be presented in the upcoming 
Sect.~\ref{sec:truckborne}.)
The features of the measurement are broadly well-reproduced by the fit,
considering the pixellization of the reconstruction.  In particular, the
magnitude, width and extent of the contamination are well represented in the
result of the fit.

\begin{figure}[h]
   \begin{center}
   \begin{tabular}{c}
   \begin{overpic}[trim = .1cm .1cm .1cm .1cm, clip = true, height=6.5cm, angle=270]{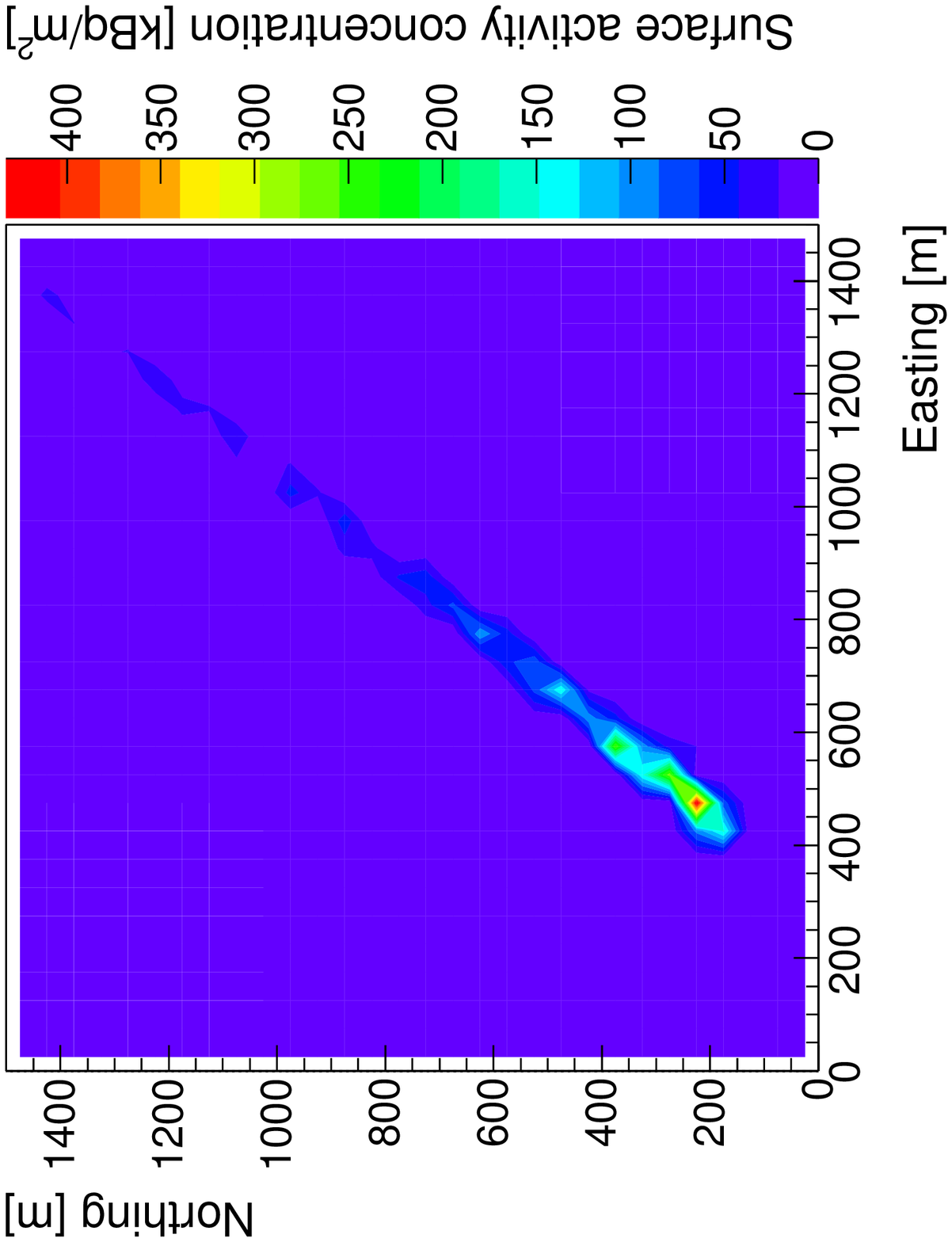}\put(18,70){\textcolor{white}{a)}}\end{overpic}\\
   \begin{overpic}[trim = .1cm .1cm .1cm .1cm, clip = true, height=6.5cm, angle=270]{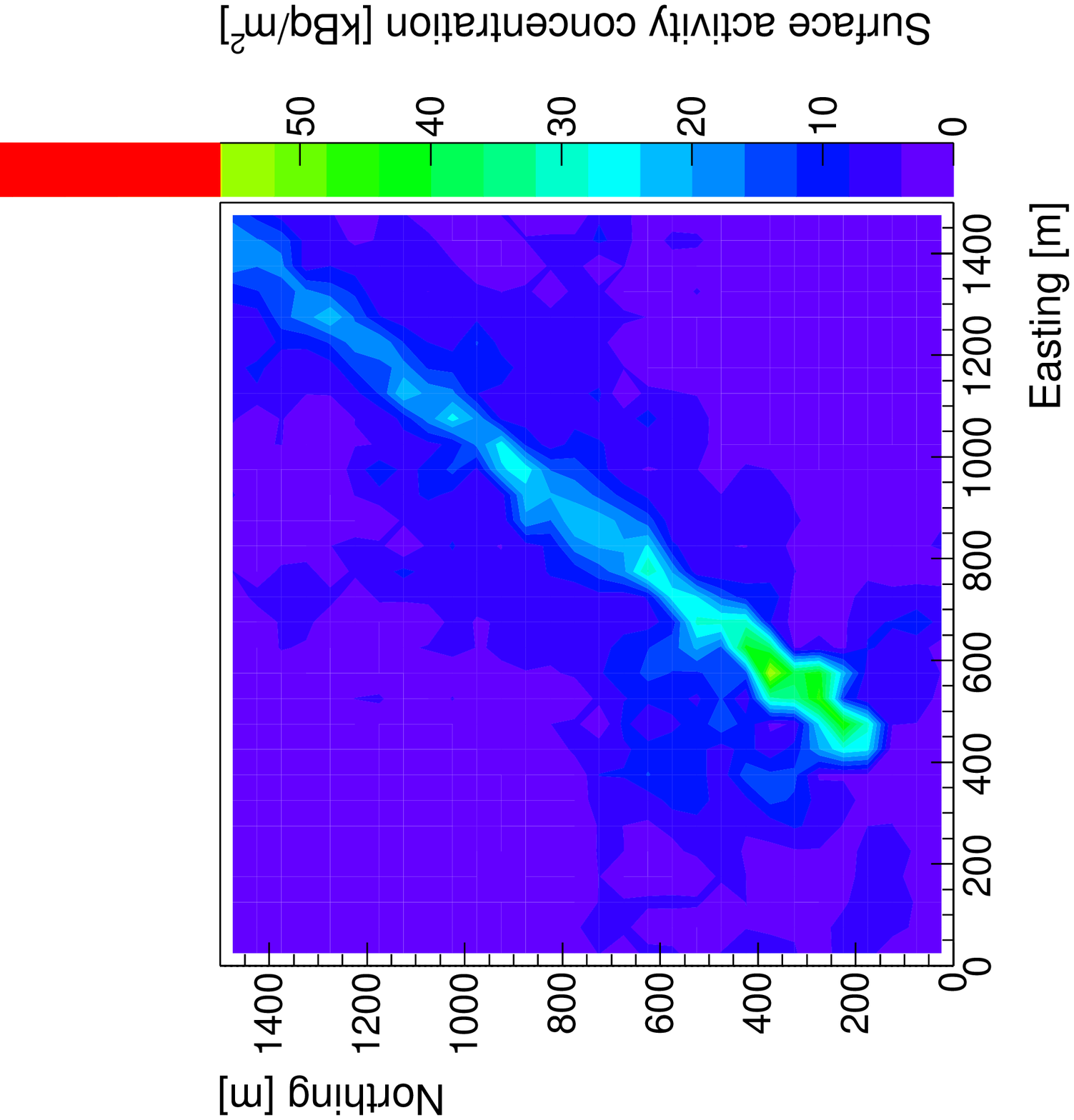}\put(18,70){\textcolor{white}{b)}}\end{overpic}
   \hspace{.4cm}
   \begin{overpic}[trim = .1cm .1cm .1cm .1cm, clip = true, height=6.5cm, angle=270]{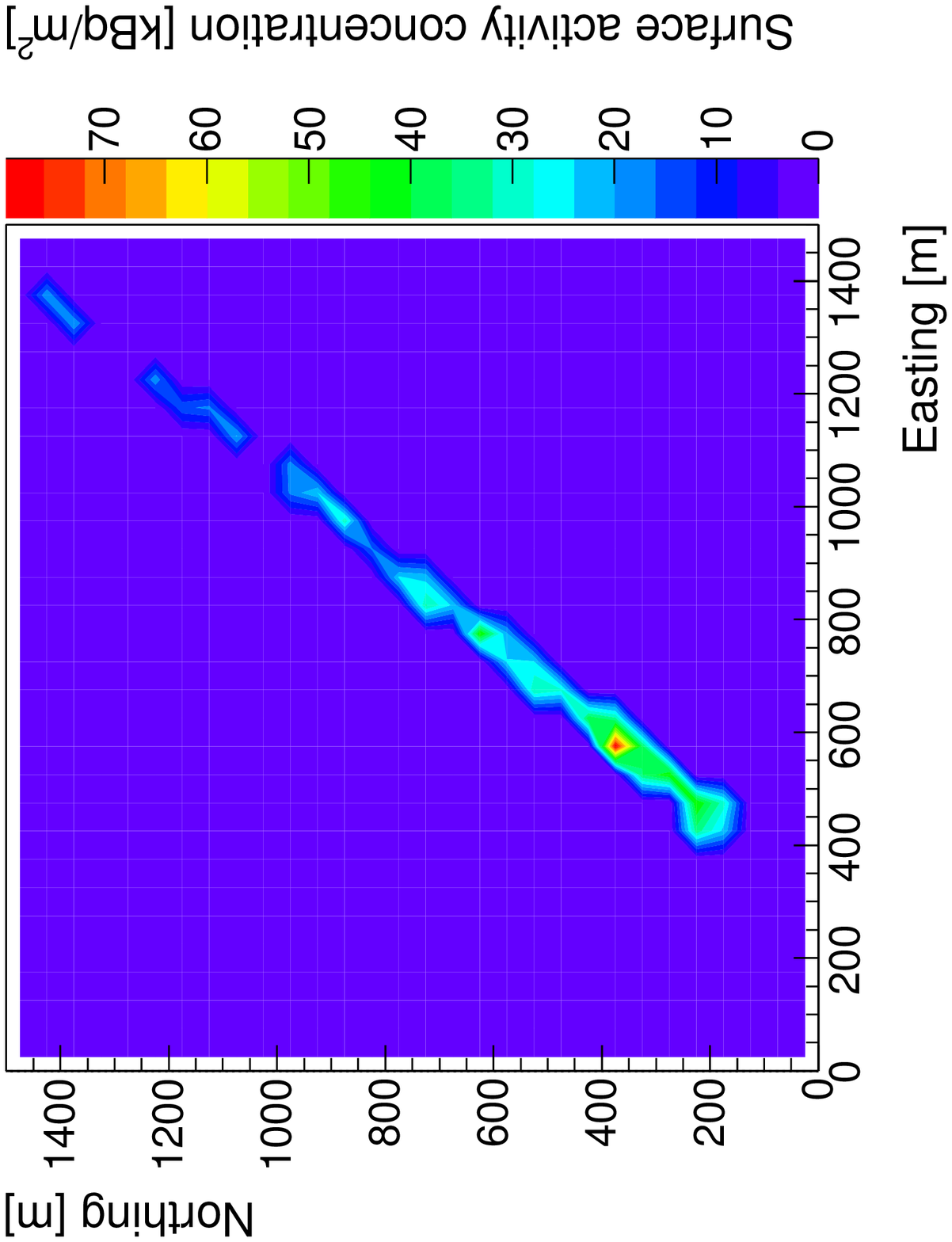}\put(18,70){\textcolor{white}{c)}}\end{overpic}\\
   \end{tabular}
   \end{center}
  \caption
  { \label{fig:data_inverted} 
a) Spatially deconvolved aerial survey measurement of fallout following
radiological dispersal device blast.
b) Positive statistical uncertainty on spatially deconvolved measurement.
c) Negative statistical uncertainty on spatially deconvolved measurement.}
\end{figure}

The underlying distribution of pixel sources which gives rise to the fit
result of Fig.~\ref{fig:data_and_fit}~b) is presented in
Fig.~\ref{fig:data_inverted}~a).
  This is the reconstructed distribution of
the surface activity concentration following spatial deconvolution.
We find that the width of the deposited plume is now much smaller than the
original undeconvolved measurement, around
50~m.  Correspondingly, the peak concentration is higher, over 400~kBq/m$^2$.
Note that the width of the deposited plume is small with respect to the altitude and line spacing of the
survey.  This accounts for its significant overestimation when the ``infinite and
uniform'' approximation was used to obtain a concentration measurement from the
measured counts as shown in Fig.~\ref{fig:data_and_fit}~a) and~\cite{Sinclair_RDD_2015}.
The length of the deposition is
however much larger than the survey parameters, so in this dimension the
``infinite and uniform'' sheet approximation is not so bad and the original
length measurement is not much altered by the spatial deconvolution.

The positive and negative statistical uncertainties on the spatially
deconvolved deposited fallout map are shown in Fig.'s~\ref{fig:data_inverted}~b)
and~c) respectively.  Note that the statistical uncertainties affecting the
measurement are very small, and on the colour scale of
Fig.~\ref{fig:data_inverted}~a) would be difficult to see, so the colour scale in
Fig.'s~\ref{fig:data_inverted}~b) and~c) is chosen to optimize the
representation of the information in Fig.~\ref{fig:data_inverted}~c).  The
uncertainties reveal important features of the measurement and its inversion.
The positive uncertainty indicates a region extending away from the measured deposited
plume axis in which a positive quantity of activity is permitted, however at a
very low amount of between 5~kBq/m$^2$ and 10~kBq/m$^2$.  The negative
uncertainty shows that the measurement of the presence and distribution of
radioactivity is significantly above zero.

The MINOS error propagation includes only the stochastic uncertainties on the
measurement.  There are additional uncertainties which are systematic and arise from
approximations in the representation of the system in the simulation.  These include mis-representation of the position, particularly the altitude; the attitude (yaw, pitch and roll); the amount of shielding material in the basket containing the detectors; the energy resolution and the air density.
The systematic uncertainty on the (undeconvolved) radioactivity concentration distribution was 
determined to be about 12\% by variation of these parameters within reasonable limits~\cite{Sinclair_RDD_2015}.

For the spatially deconvolved measurement, some of the systematic
uncertainties must be re-examined as they can be expected to have an effect on the shape of the reconstructed fallout distribution, as well as its overall magnitude.  These are the systematic uncertainties associated with the measurement of altitude and pitch angle.  
It is also interesting to examine the effect of the measurement of yaw angle
on the spatially inverted measurement as it can have no effect at all on the
original undeconvolved measurement which used the infinite sheet approximation
for the source and therefore the detector response was invariant under changes of yaw.
 
The inertial navigation system
determined the yaw angle 
during the measurement to be around -30$^\circ$.  The spatial inversion was conducted
twice.  For the central value of the inversion as presented in 
Fig.~\ref{fig:data_inverted}~a) the helicopter systems in the template
histograms were assigned a yaw of -30$^\circ$ to match the data.  To allow for
changes in yaw during flight, the regression was repeated with yaw set
maximally different at 60$^\circ$.  Pitch was varied from the nominal
0$^\circ$ to -10$^\circ$ according to the maximum deviation of pitch recorded
by the inertial navigation system while on line during one sortie.
Altitude was varied from nominal by 1~m to account for approximately one sigma
of variability in height determination.
These variations did not significantly alter the measurement of length and
width of the deposited fallout.
Added in quadrature, and considering that some of the variation was already
included in the original systematic uncertainty associated with the
sensitivity, the deviations do not yield a significant additional systematic
uncertainty.
Although not a significant additional source of uncertainty for the measurements
presented here, these sources of uncertainty are worth discussing for the
benefit of researchers following this approach under different operating conditions.

\subsubsection{Comparison of spatially deconvolved aerial survey data
  with truckborne survey data}
\label{sec:truckborne}

\begin{figure}[h]
   \begin{center}
   \begin{tabular}{c}
   \begin{overpic}[trim = .1cm .1cm .1cm .1cm, clip = true, height=6.5cm, angle=270]{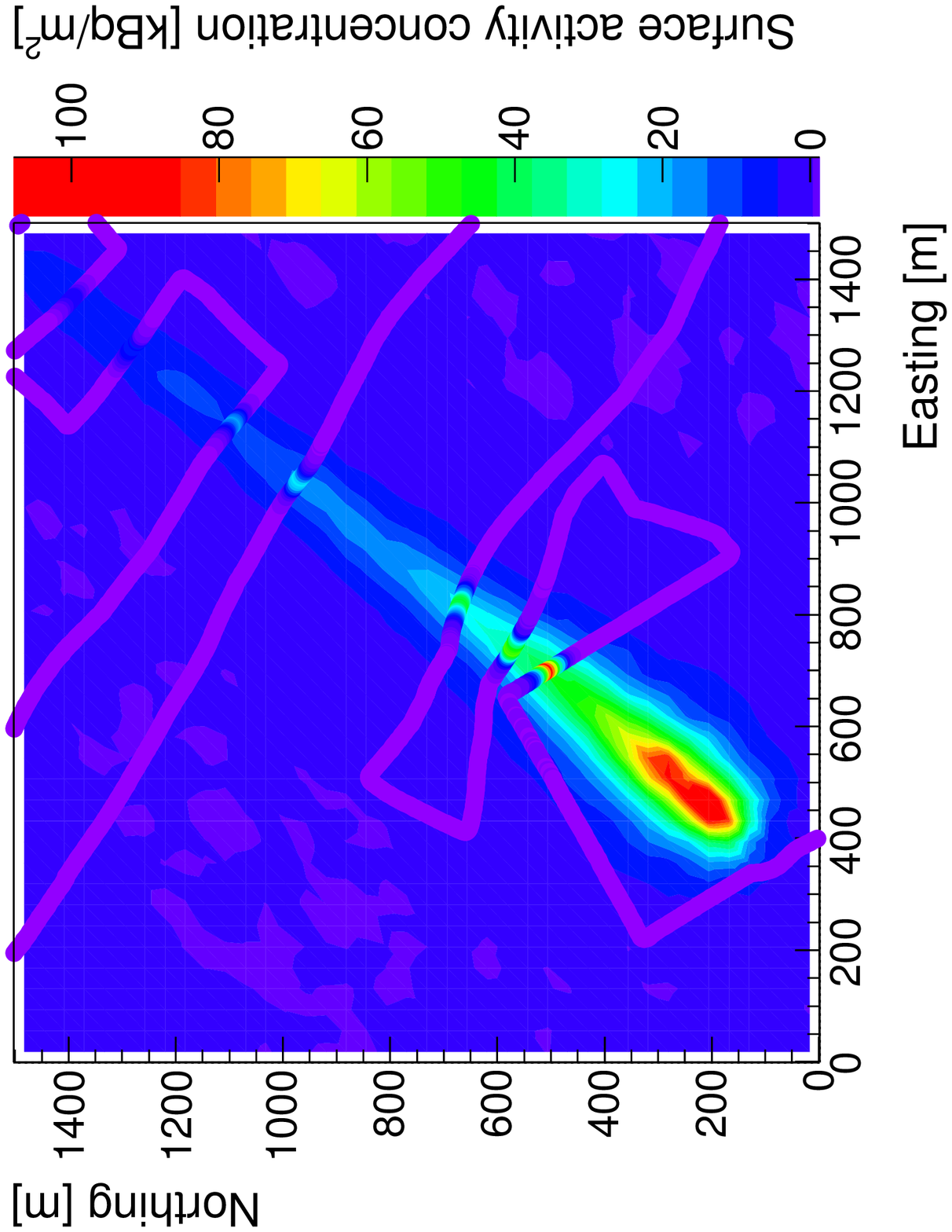}\put(18,70){\textcolor{white}{a)}}\end{overpic}
   \hspace{.4cm}
   \begin{overpic}[trim = .1cm .1cm .1cm .1cm, clip = true, height=6.5cm, angle=270]{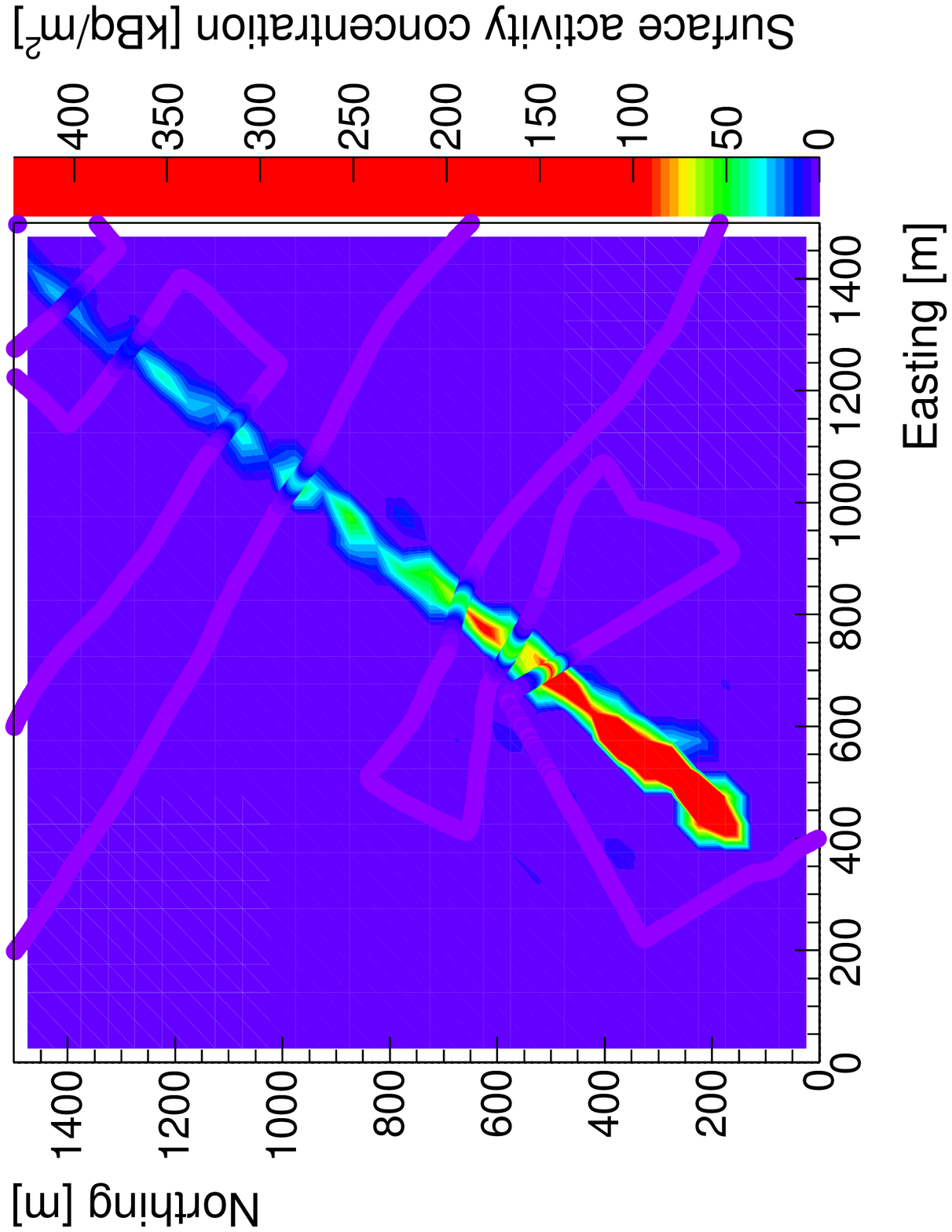}\put(18,70){\textcolor{white}{b)}}\end{overpic}\\
   \vspace*{.5cm}\\   
   \begin{overpic}[height=6.5cm, angle=0]{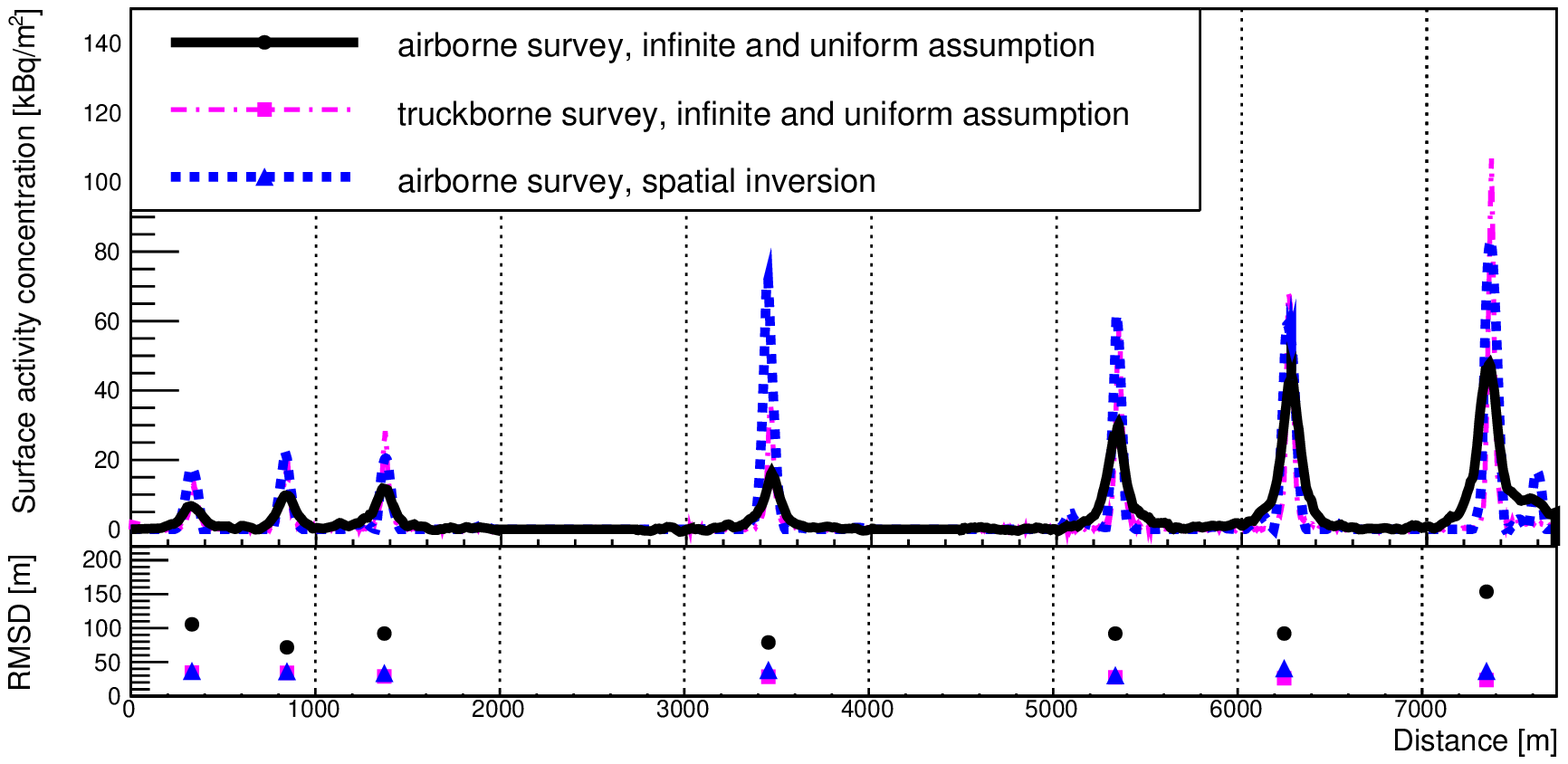}\put(95,45){\textcolor{black}{c)}}\end{overpic}
   \end{tabular}
   \end{center}
  \caption
  { \label{fig:cftruck} 
a) Aerial radiation survey following detonation of radiological dispersal
device, with results of radiation survey from a truck-based system overlaid.
b) Spatially deconvolved aerial radiation survey of the RDD fallout, with
results of radiation survey from a truck-based system.  Colour scale is the
same in a) and b) and is optimized to show the range of values of the result from the truckborne survey.
c) (top) Transects of the deposited RDD plume following the path of the truck-based survey
system.  The
solid line shows the aerial survey result sampled at the truck location.  The dot-dashed line shows the
truckborne survey result and the dashed line shows the aerial survey result
after spatial deconvolution sampled at the truck location. (bottom)
Root-mean-square deviation of the seven transects versus the location of maximum
concentration of each transect according to the truckborne survey.  Circles
show the aerial survey result, squares show the truckborne survey and
triangles show the spatially deconvolved aerial survey result.}
\end{figure}

In Fig.~\ref{fig:cftruck}a) the truckborne survey result which followed the
first blast is shown overlaid on
the undeconvolved aerial survey result from the same blast
on the same colour scale.  
(The colour
scale is optimized to show the variation in the truckborne survey result.)
The truckborne survey reports much higher surface activity oncentrations than
the aerial survey and the fallout appears to be narrower in width.

Fig.~\ref{fig:cftruck}b) shows the same truckborne survey result this time
overlaid on the spatially deconvolved aerial survey measurement.  (The colour
scale is the same as that used in Fig.~\ref{fig:cftruck}a).)
Here, both the width and magnitude of the concentration are in better
agreement.

The aerial survey maps were sampled at the locations of the truck and these
sampled activity concentration values are presented in
Fig.~\ref{fig:cftruck}c) (top) as a function of the distance from the start of the
truck-driven sortie.
Again, it is clear that the spatially deconvolved result is generally higher
at maximum
magnitude and more narrow than the undeconvolved aerial survey result.  
The bottom part of Fig.~\ref{fig:cftruck}c) shows the root-mean-square
deviation (RMSD)
of the curves calculated by sampling each profile at regular intervals
from the maximum concentration down to the point at which the concentration
falls below 10\% of the maximum.  This plot shows that the RMSD width
of the deposition according to the undeconvolved survey is typically around
90~m while the width of the deposition according to the
truckborne survey and the spatially deconvolved aerial survey tends to be
significantly narrower, closer to 30~m.
Spatially deconvolved aerial survey thus recovers the
narrowness of the fallout to approximately the same spectral
precision as the truckborne survey.

Truckborne data is shown for the purpose of shape comparison only and does not
include detailed error analysis.  In any event, the truckborne system has its
own finite area of sensitivity largely caused by its ``altitude'' of just over one
metre, causing smearing of the measured spatial distribution.  
A contact measurement of the deposited radioactivity can be expected to be even more
concentrated in places~\cite{Erhardt_deposition_RDD_2015}.

\section{Discussion and Conclusions}

Radiometric survey would be performed to map fallout following a reactor
accident or following a malicious
release.  To cover a large area quickly, the surveys are initially performed
using manned aircraft at some significant altitude $H$.  If there is spatial
variability in the fallout at distance scales much less than $H$, then
the map result of the survey can underestimate the quantitity of
radioactivity on the ground in places.

We have presented here a method to deconvolve an aerial survey map for the
spatial smearing caused by measurement at altitude, at least to the extent
permitted by the sampling density as determined by the aircraft speed and line
spacing.
Performed on synthetic data, the deconvolution method returns a distribution
which is consistent with the true underlying distribution within
measurement uncertainty.
The deconvolved distribution is more narrowly distributed, and shows regions
of locally higher radioactivity concentration than the initial undeconvolved
measurement.
Performed on real data acquired following detonation of a radiological
dispersal device, the method produces a distribution which is narrower and
shows radioactivity concentration as much as four times that of the original
measurement.

The method can unfold a distribution for smearing effects up to a resolution
permittable by the sampling frequency of the original measurement.  The
method allows for propagation of stochastic measurement
uncertainties through the unfolding to obtain the measurement uncertainties on
the fit parameters.
The method relies on application of the MINUIT and MINOS algorithms well known
in the field of particle physics.
What is perhaps not well known is that these algorithms can tolerate
operating with hundreds of independent fit parameters, converging to a stable solution in
a reasonable amount of time from an arbitrary starting distribution.
Our current need was to develop a method to extract the greatest possible
information from a set of aerial surveys performed to improve scientific
understanding of the behaviour of radiological dispersal devices.
The method, however, is also applicable to unfolding of any smeared distribution
of any dimensionality.  It could find application in other fields.

The result of the unfolding is limited in spatial resolution by the requirement
that the density of pixellization of the answer not exceed the density of
measurements as determined by the aircraft speed and line spacing.  
Nevertheless, aerial survey practitioners should be aware that there is 
improved information about the spatial distribution of the radioactivity 
contained in their aerial survey map that can be extracted provided good 
knowledge of the response function of the system is available.
The achieved spatial resolution for the particular aerial survey following RDD 
detonation presented here approximately matched that obtained during
truckborne survey over the same deposition (while providing complete coverage).
Contact measurements ($H=0$~m) can be expected to reveal even greater
local spatial variations than the truckborne data.
Still, the truckborne survey ``height'' of about 1.2~m provides a salient
benchmark for spatial resolution as this is close to the average height of an
adult human.
Should humans be required to enter a possibly contaminated area guided by the
results of aerial survey alone, a spatially deconvolved aerial survey map could
provide a better predictor of the activity concentrations they will encounter
than the undeconvolved measurement.

\section*{Acknowledgements}

The authors gratefully acknowledge the leadership of the RDD field trials
under L. Erhardt, and helpful comments on the analysis from H.C.J.~Seywerd,
P.R.B.~Saull and F.A,~Marshall.  Funding for this project was provided through
Canada's Chemical, Biological, Radiological-Nuclear and Explosives Research
and Technology Initiative.  This report is NRCan Contribution 20180112.





\bibliographystyle{elsarticle-num}
\bibliography{NIM_2015}







\end{document}